\documentclass[10pt]{article}

\usepackage{color}
\usepackage{gensymb}
\usepackage{upgreek}

\usepackage{changepage}

\usepackage[utf8]{inputenc}

\usepackage{textcomp,marvosym}

\usepackage{fixltx2e}

\usepackage{amsmath,amssymb}

\usepackage{gensymb}

\usepackage{cite}

\usepackage{nameref,hyperref}

\usepackage[right]{lineno}

\usepackage{microtype}
\DisableLigatures[f]{encoding = *, family = * }

\usepackage{rotating}

\PassOptionsToPackage{version=3}{mhchem}
\usepackage{mhchem}

\usepackage[aboveskip=1pt,labelfont=bf,labelsep=period,justification=raggedright,singlelinecheck=off]{caption}

\makeatletter
\renewcommand{\@biblabel}[1]{\quad#1.}
\makeatother

\date{}

\global\long\def\ddt{\frac{\mathrm{d}}{\mathrm{d}t}}

\begin{document}
\vspace*{0.35in}

\begin{flushleft}
{\Large
\textbf\newline{How to distinguish conformational selection and induced fit based on chemical relaxation rates} 
}
\newline
\\
Fabian Paul\textsuperscript{1,2},
Thomas R.\ Weikl\textsuperscript{1}
\\
\bigskip\small
1 Max Planck Institute of Colloids and Interfaces, Department of Theory and Bio-Systems, Science Park Golm, 14424 Potsdam, Germany
\\
2 Free University Berlin, Department of Mathematics and Computer Science, Arnimallee 6, 14195 Berlin, Germany
\bigskip

%
%






\end{flushleft}


\section*{Abstract}
Protein binding often involves conformational changes. Important questions are whether a conformational change occurs prior to a binding event (`conformational selection') or after a binding event (`induced fit'), and how conformational transition rates can be obtained from experiments. In this article, we present general results for the chemical relaxation rates of conformational-selection and induced-fit binding processes that hold for all concentrations of proteins and ligands and, thus, go beyond the standard pseudo-first-order approximation of large ligand concentration. These results allow to distinguish conformational-selection from induced-fit processes -- also in cases in which such a distinction is not possible under pseudo-first-order conditions -- and to extract conformational transition rates of proteins from chemical relaxation data. 

\section*{Introduction}

\begin{figure*}[t]
\begin{center}
\hspace*{-2.cm}
\resizebox{1.3\columnwidth}{!}{\includegraphics{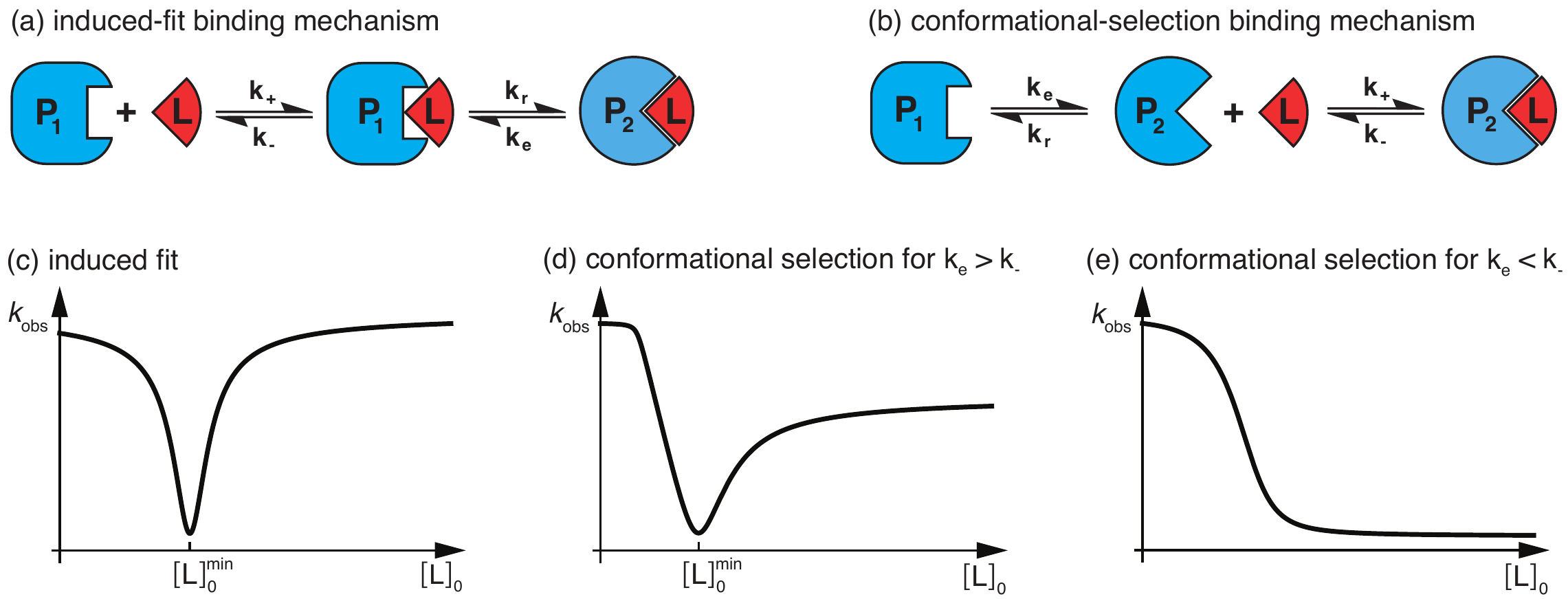}}
\end{center}
\caption{\small{\bf Characteristic chemical relaxation of induced-fit and conformational-selection binding}. (a) In induced-fit binding, the change between the conformations $\mathrm{P}_1$ and $\mathrm{P}_2$ of the protein occurs after binding of the ligand L. The intermediate state $\mathrm{P}_1\mathrm{L}$ relaxes into the bound ground state $\mathrm{P}_2\mathrm{L}$  with rate $k_r$, and is excited from the  ground state with rate $k_e$. (b) In conformational-selection binding, the conformational change of the protein occurs prior to ligand binding. The intermediate state $\mathrm{P}_2$ is excited from the unbound ground state $\mathrm{P}_1$ with rate $k_e$, and relaxes back into the ground state with rate $k_r$. (c) The dominant, smallest relaxation rate $k_\text{obs}$ of induced-fit binding is minimal at the total ligand concentration $[\text{L}]_0^\text{min} = [\text{P}]_0 - K_d$ where $[\text{P}]_0$ is the total protein concentration and $K_d$ the overall dissociation constant. As a function of $[\text{L}]_0$, the dominant rate $k_\text{obs}$ is symmetric with respect to this minimum. (d) The dominant, smallest relaxation rate $k_\text{obs}$ of conformational-selection binding has a characteristic minimum as a function of $[\text{L}]_0$ for $k_e > k_{-}$ , but is not symmetric with respect to this minimum. (e)  The dominant rate $k_\text{obs}$ of conformational-selection binding decreases monotonically with $[\text{L}]_0$ for $k_e < k_{-}$.
}
\label{figure_general}
\end{figure*}

Protein function often involves conformational changes during the binding to ligand molecules \cite{Gerstein98}. Advanced NMR experiments \cite{Eisenmesser05,Beach05,Boehr06a,Henzler07b,Tang07,Lange08} and single-molecule spectroscopy \cite{Kim13,Munro14,Ghoneim14} indicate that these conformational changes can occur without ligand, or with bound ligand and thus point to an intrinsic conformational dynamics of the proteins. An important question is how the conformational dynamics is coupled to the binding events. Two mechanisms for this coupling are `conformational selection' \cite{Ma99} and `induced fit' \cite{Koshland58} (see Fig.\ \ref{figure_general}(a) and (b)). In conformational-selection binding, a conformational change occurs {\em prior to} the binding of a ligand molecule, as a conformational excitation from the unbound-ground state conformation of the protein. In this mechanism, the ligand seems to `select' and stabilize a higher-energy conformation for binding. In induced-fit binding, the conformational change occurs {\em after} ligand binding and is a conformational relaxation into the bound ground-state conformation that is apparently `induced' by the ligand. These two mechanisms are in particular plausible for small ligand molecules that can quickly `hop' in and out of the protein binding pocket, i.e.\ that can enter and exit this binding pocket within transition times that are significantly smaller than the residence or dwell times of the proteins in the different conformations \cite{Weikl14}. 
   
A central problem is to identify protein binding mechanisms based on experimental data \cite{Bosshard01,Sullivan08,Weikl09,Boehr09,Hammes09,Wlodarski09,Changeux11b,Weikl12,Vogt12,Kiefhaber12,Weikl14,Vogt14}. Advanced NMR experiments and single-molecule spectroscopy can reveal higher-energy conformations that are necessary for conformational-selection or induced-fit binding, but do not directly indicate the binding mechanism because such higher-energy conformations may exist both in the bound and unbound state of the protein \cite{Boehr06a,Kim13}. In principle, both conformational-selection or induced-fit binding then are possible. Standard mixing or temperature-jump experiments that probe the chemical relaxation into the binding equilibrium can provide additional information that allows to identify the binding mechanism\cite{Pozzi12,Vogt12,Daniels14,Daniels15,Chakrabarti16}. Of particular interest is the dominant, slowest relaxation rate $k_\text{obs}$ observed in the experiments, and how this rate depends on the total ligand concentration $[\text{L}]_0$ \cite{Pozzi12,Vogt12,Chakrabarti16}. The chemical relaxation experiments are often performed and analysed under pseudo-first-oder conditions, i.e.\ at ligand concentrations that greatly exceed the protein concentrations \cite{James03,Heredia06,Kim07,Tummino08,Antoine09,Pozzi12,Vogt12,Vogt13,Gianni14,Vogt15}. In the case of  induced-fit binding, the dominant relaxation  rate $k_\text{obs}$ {\em increases} monotonically with the ligand concentration $[\text{L}]_0$ under pseudo-first-oder conditions. In the case of conformational-selection binding, $k_\text{obs}$ {\em decreases} monotonically with increasing $[\text{L}]_0$ for conformational excitation rates $k_e<k_{-}$, and {\em increases} monotonically with $[\text{L}]_0$ for $k_e>k_{-}$ where $k_{-}$ is the unbinding rate of the ligand from the bound ground-state conformation of the protein (see Fig.\ \ref{figure_general}(b)). A decrease of the dominant relaxation rate $k_\text{obs}$ with increasing ligand concentration $[\text{L}]_0$ thus indicates conformational-selection binding \cite{Pozzi12}. However, an increase of $k_\text{obs}$ with $[\text{L}]_0$ under pseudo-first-order conditions is possible both for induced-fit binding and conformational-selection binding and does not uniquely point towards a binding mechanism \cite{Vogt12}. 

In this article, we present general analytical results for the dominant relaxation rate $k_\text{obs}$ of induced-fit binding and conformational-selection binding processes that hold for all ligand and protein concentrations. Our general results are based on an expansion of the rate equations for these binding processes around the equilibrium concentrations of ligands and proteins, and include the pseudo-first-order results in the limit of large ligand concentrations. In the case of induced-fit binding, we find that $k_\text{obs}$ exhibits a minimum at the total ligand concentration $[\text{L}]_0^\text{min} = [\text{P}]_0 - K_d$ for total protein concentrations $[\text{P}]_0$ that are larger than the overall dissociation constant $K_d$ of the binding process. As a characteristic feature, the function $k_\text{obs} ([\text{L}]_0)$ for induced-fit binding is symmetric with respect to this minimum. At sufficiently large protein concentrations $[\text{P}]_0$, the function $k_\text{obs} ([\text{L}]_0)$ tends to identical values for small ligand concentrations $[\text{L}]_0 \ll [\text{P}]_0$ and for large ligand concentrations $[\text{L}]_0 \gg [\text{P}]_0$ because of its symmetry (see Fig.\ \ref{figure_general}(c)).   In the case of conformational-selection binding, we find that $k_\text{obs}$ exhibits a minimum for conformational excitation rates $k_e>k_{-}$ and sufficiently large protein concentrations $[\text{P}]_0$ (see Fig.\ \ref{figure_general}(d)). The location $[\text{L}]_0^\text{min}$ of this minimum depends on $[\text{P}]_0$,  $K_d$,  and the rates $k_e$ and $k_{-}$ (see Eq.\ (\ref{L0min_CS}) below). In contrast to induced-fit binding, the function $k_\text{obs} ([\text{L}]_0)$ for conformational-selection binding is not symmetric with respect to this minimum. At sufficiently large protein concentrations $[\text{P}]_0$, the function $k_\text{obs} ([\text{L}]_0)$ attains values for small ligand concentrations $[\text{L}]_0 \ll [\text{P}]_0$ that can greatly exceed the values for large ligand concentrations  $[\text{L}]_0 \gg [\text{P}]_0$ (see Fig.\ \ref{figure_general}(d)). For excitation rates $k_e<k_{-}$ of conformational-selection binding processes, the dominant relaxation rate $k_\text{obs}$ decreases monotonically  with increasing ligand concentration $[\text{L}]_0$ (see Fig.\ \ref{figure_general}(e)). Our general results for the dominant relaxation rate $k_\text{obs}$ of induced-fit and conformational-selection binding processes allow to clearly distinguish between these two binding mechanisms for sufficiently large protein concentrations $[\text{P}]_0$ (see Figs. \ref{figure-example-CS} and  \ref{figure-example-IF} below for numerical examples).

%
 
\section*{Results}

Solving the rate equations of the induced-fit and conformational-selection binding models shown in Fig.\ \ref{figure_general}(a) and (b) is complicated by the fact that the binding steps in these models are second-order reactions that depend on the product of the time-dependent concentrations of unbound proteins and unbound ligands. In the standard pseudo-first-order approximation,  the rate equations are simplified by assuming that the total ligand concentration greatly exceeds  the total protein concentration, so that the  amount of ligand consumed during binding is negligible compared to the total amount of ligand. The concentration of the unbound ligand then can be taken to be constant, and the rate equations only contain terms that are linear in the time-dependent concentration of the protein, which makes them solvable. In our more general approach, a linearization of the rate equations is achieved by expanding around the equilibrium concentrations of the bound and unbound proteins and ligands (see Methods). This expansion captures the final relaxation into equilibrium, which is governed by the smallest, dominant relaxation rate $k_\text{obs}$, for all concentrations of proteins and ligands, and leads to general results for $k_\text{obs}$ that include the results from the pseudo-first-order approximation in the limit of large ligand concentrations. 

\subsection*{Dominant relaxation rate of induced-fit binding}

Expanding the rate equations of the induced-fit binding mechanism shown in Fig.\ \ref{figure_general}(a) around the equilibrium concentrations of proteins and ligands leads to the dominant, smallest relaxation rate (see Methods)
\begin{equation}
k_\text{obs}=k_e+k_r+\frac{1}{2}\gamma - \frac{1}{2}\sqrt{\gamma^{2} + 4k_{-}k_r }
\label{if-kobs}
\end{equation}
with
\begin{align}
&\gamma= -k_e-k_r+k_{-}+k_{+}\left(\delta-K_d\right)
\label{if-gamma} \\
&\delta=\sqrt{\left([\text{L}]_{0}- [\text{P}]_{0}+K_d\right)^2+4[\text{P}]_{0}K_d}
\label{delta}
\end{align}
and with the overall dissociation constant
\begin{equation}
K_d = \frac{k_{-}k_e}{k_{+}(k_e+ k_r)}
\label{if-Kd}
\end{equation}
of induced-fit binding. This general result for $k_\text{obs}$ holds for all total ligand concentrations $[\text{L}]_{0}$ and protein concentrations $[\text{P}]_{0}$. In the limit of large ligand concentrations $[\text{L}]_{0} \gg [\text{P}]_{0}$, we obtain $\delta \simeq [\text{L}]_{0} +K_d$ and $\gamma \simeq -k_e-k_r+k_{-}+k_{+}[\mathrm{L}]_{0}$ from Eqs.\ (\ref{if-gamma}) and (\ref{delta}), which agrees with results derived in pseudo-first-order approximation \cite{Weikl12,Vogt12}. 

As a function of the total ligand concentration $[\text{L}]_{0}$, the dominant relaxation rate $k_\text{obs}$ exhibits a minimum at 
\begin{equation}
[\text{L}]_0^\text{min} = [P]_0 - K_d
\label{if-L0min}
\end{equation}
for total protein concentrations $[P]_0 > K_d$. The function $k_\text{obs}([\text{L}]_{0})$ is symmetric with respect to $[\text{L}]_0^\text{min}$ (see Fig.\ (\ref{figure_general}(c)). This symmetry and the location $[\text{L}]_0^\text{min}$ of the minimum result from the fact that $k_\text{obs}$ depends on $[\text{L}]_{0}$ only via the term $\delta$, which is minimal at $[\text{L}]_0^\text{min}$ and symmetric with respect to $[\text{L}]_0^\text{min}$. The dominant relaxation rate $k_\text{obs}$ is minimal when $\delta$ is minimal. 
For large ligand concentrations $[\text{L}]_{0}$, $k_\text{obs}$ tends towards the maximum value $k_e + k_r$ as in pseudo-first-order approximation. The location $[\text{L}]_0^\text{min}$ of the minimum and the symmetry of the function $k_\text{obs}([\text{L}]_{0})$ with respect to this minimum are properties that the induced-fit binding model appears to `inherit' from the elementary binding model $\ce{P + L <=> PL}$ (see Eq.\ (\ref{eb-kobs}) in Methods section). However, the function $k_\text{obs}([\text{L}]_{0})$ of the elementary binding model is V-shaped and does not tend to a constant maximum value for large  ligand concentrations $[\text{L}]_{0}$. 

\subsection*{Dominant relaxation rate of conformational-selection binding}

For the conformational-selection binding mechanism shown in Fig.\ \ref{figure_general}(b),  an expansion of the rate equations around the equilibrium concentrations of proteins and ligands leads to the dominant, smallest relaxation rate (see Methods)
\begin{equation}
k_\text{obs}=k_e+\frac{1}{2}\alpha - \frac{1}{2}\sqrt{\alpha^{2}+\beta}
\label{cs-kobs}
\end{equation}
with
\begin{align}
&\alpha =  k_r - k_e + \frac{k_{-}\left((2 k_e + k_r)\delta  + k_r \left( [L]_0 - [P]_0 - K_d\right)\right)}{2 k_e K_d} 
\label{cs-alpha}
\\
&\beta =  2 k_r \left(2 k_e - k_{-} - \frac{k_{-}\left(\delta - [L]_0 + [P]_0 \right) }{K_d} \right)
\label{cs-beta}
\end{align}
and $\delta$ as in Eq.\ (\ref{delta}), and with the overall dissociation constant 
\begin{equation}
K_d = \frac{k_{-}(k_e + k_r)}{k_{+}k_e}
\label{cs-Kd}
\end{equation}
of conformational-selection binding. This general result for $k_\text{obs}$ holds for all total ligand concentrations $[\text{L}]_{0}$ and protein concentrations $[\text{P}]_{0}$. In the limit of large ligand concentrations $[\text{L}]_{0} \gg [\text{P}]_{0}$, we obtain $\alpha\simeq -k_e+k_r+k_{-}+k_{+}[\mathrm{L}]_{0}$ and $\beta \simeq 4k_r(k_e-k_{-})$ from Eqs.\ (\ref{delta}), (\ref{cs-alpha}), and (\ref{cs-beta}), in agreement with results derived in pseudo-first-order approximation \cite{Weikl12,Vogt12}. 

For conformational-selection binding, the shape of the function $k_\text{obs}([\text{L}]_{0})$  depends on the values of the conformational excitation rate $k_e$ and the unbinding rate $k_{-}$ (see Fig.\ \ref{figure_general}(d) and (e)).  For $k_e < k_{-}$,  the dominant relaxation rate $k_\text{obs}$ decreases monotonically with increasing total ligand concentration $[\text{L}]_{0}$. For $k_e > k_{-}$, the dominant relaxation rate $k_\text{obs}$ exhibits a minimum as a function of $[\text{L}]_{0}$ at sufficiently large total protein concentrations $[\text{P}]_0$. The minimum is located at (see Methods)
\begin{equation}
[\text{L}]_{0}^{\min}\simeq\frac{k_e+k_{-}}{k_e-k_{-}}[\text{P}]_{0}-K_d
\label{L0min_CS}
\end{equation}
if the conformational relaxation rate $k_r$ is much larger than the excitation rate $k_e$, which typically holds for the conformational exchange between ground-state and excited-state conformations of proteins. In contrast to induced-fit binding, the function $k_\text{obs}([\text{L}]_{0})$ is not symmetric with respect to this minimum. For large ligand concentrations, the limiting value of the dominant relaxation rate is $k_\text{obs}(\infty) = k_e$ as in pseudo-first-order approximation.  For vanishing ligand concentrations $[\text{L}]_0 \to 0$,  the limiting value is $k_\text{obs}(0) = k_e + k_r$ for total protein concentrations $[\text{P}]_0 > K_d(k_e + k_r -k_{-})/k_{-}$ and $k_\text{obs}(0) = k_{-}([\text{P}]_0 + K_d)/K_d$ for $[\text{P}]_0<K_d(k_e + k_r -k_{-})/k_{-}$.

\subsection*{Distinguishing induced fit and conformational selection}
\begin{figure*}[t]
\begin{center}
\hspace*{-2cm}
\resizebox{1.3\columnwidth}{!}{\includegraphics{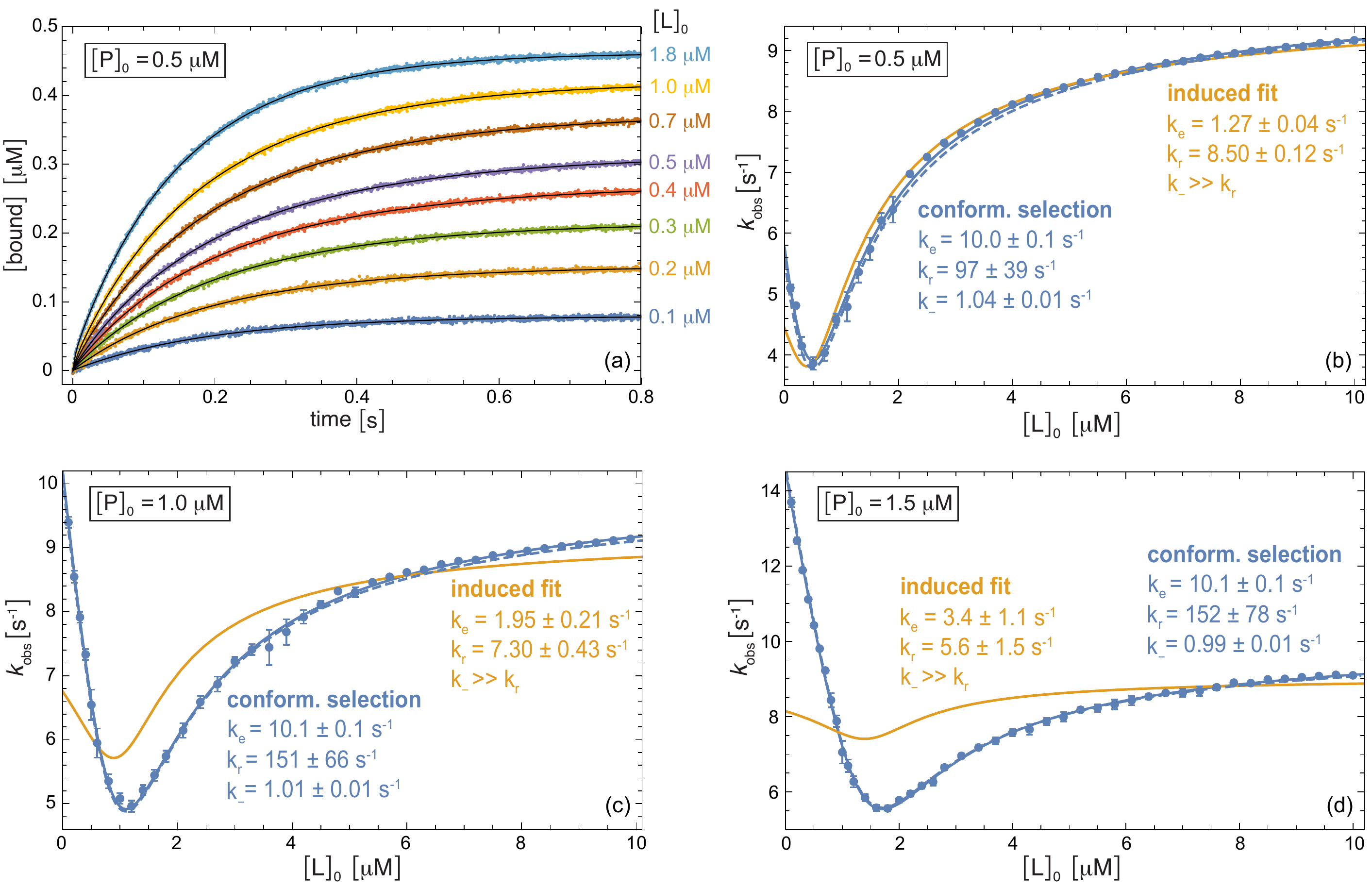}}
\end{center}
\caption{\small {\bf Numerical example for conformational-selection binding} with the rate constants $k_e = 10\; \text{s}^{-1}$, $k_r = 100\; \text{s}^{-1}$, $k_{+} = 100 \; \upmu\text{M}^{-1} \text{s}^{-1}$, and $k_{-} = 1\; \text{s}^{-1}$. (a) Relaxation data for the bound complex obtained by numerical integration of the rate equations and subsequent addition of Gaussian noise with amplitude $0.002 \; \upmu\text{M}$ at the total protein concentration $[\text{P}]_0 = 0.5\; \upmu$M and exemplary total ligand concentrations $[\text{L}]_0$. The black lines represent multi-exponential fits of the data points. (b) to (d) Comparison of $k_\text{obs}$ values obtained from multi-exponential fits of numerical relaxation data (points) to our theoretical results for $k_\text{obs}$ (lines) at the three different total protein concentrations $[\text{P}]_0 = 0.5\, \upmu\text{M}$, $1.0\; \upmu\text{M}$, and $1.5\; \upmu\text{M}$ and total ligand concentrations $[\text{L}]_0$ between $0.1\, \upmu\text{M}$ and $10\, \upmu\text{M}$. The full lines represent fits of Eq.\ (\ref{cs-kobs}) for conformational-selection binding (blue) and of Eq.\ (\ref{if-kobs}) for induced-fit binding (orange), with fit parameter values specified in the figure. In these fits, the dissociation constant $K_d = 0.11 \; \upmu\text{M}$ is assumed to be known from equilibrium data. The dashed blue lines are obtained from Eq.\ (\ref{cs-kobs}) for the `true' rate constants of the numerical example. 
}
\label{figure-example-CS}
\end{figure*}

The general results for the dominant relaxation rate $k_\text{obs}$ presented in the previous sections allow to clearly distinguish induced-fit from conformational-selection binding processes. In Fig.\ \ref{figure-example-CS}, we consider a conformational-selection binding process with the rate constants $k_e = 10\; \text{s}^{-1}$, $k_r = 100\; \text{s}^{-1}$, $k_{+} = 100 \; \upmu\text{M}^{-1} \text{s}^{-1}$, and $k_{-} = 1\; \text{s}^{-1}$ as a numerical example. The data points in Fig.\ \ref{figure-example-CS}(a) represent relaxation curves for the bound complex that have been generated by numerical integration of the rate equations and subsequent addition of Gaussian noise to mimic measurement errors. The black lines in Fig.\ \ref{figure-example-CS}(a) are multi-exponential fits of the data points. The number of exponentials in these fits has been determined with the Akaike information criterion (AIC), which is a standard criterion for the trade-off between quality of fit and number of fit parameters, and ranges from 2 to 4. The data points in Fig.\ \ref{figure-example-CS}(b) to (d) represent the dominant relaxation rates $k_\text{obs}$ that are obtained from multi-exponential fits of relaxation curves for different total ligand concentrations $[\mathrm{L}]_0$ and total protein concentrations $[\mathrm{P}]_0$. The dominant relaxation rate $k_\text{obs}$ here is identified as the smallest relaxation rate of a multi-exponential fit. The full blue lines in  Fig.\ \ref{figure-example-CS}(b) to (d) result from fitting our general result (\ref{cs-kobs}) for conformational-selection binding to the $k_\text{obs}$ data points. The full orange lines represent fits of our general result (\ref{if-kobs})  for induced-fit binding. For all fits, we assume that the dissociation constant $K_d = 0.11 \; \upmu\text{M}$ is known from equilibrium data, and use $k_e$, $k_r$, and $k_{-}$ as fit parameters. Finally, the blue dashed lines in Figs.\ \ref{figure-example-CS}(b) to (d) are the $k_\text{obs}$ curves obtained from Eq.\ (\ref{cs-kobs}) for the `true' rate constants of the conformational-selection binding process given above. These dashed lines agree with the data points, which indicates that the $k_\text{obs}$ values from multi-exponential fits as in Fig.\ \ref{figure-example-CS}(a) are identical to the values obtained from Eq.\ (\ref{cs-kobs}) within the statistical errors of the numerical example. 
 
The fits in Fig.\ \ref{figure-example-CS}(b) to (d) clearly identify conformational selection as the correct binding mechanism in this example. The blue fit curves for conformational selection agree with the data points within statistical errors, while the orange fit curves for induced fit deviate from the data. For conformational-selection binding, the fit values of the conformational transition rates $k_e$ and $k_r$ and of the unbinding rate $k_{-}$ specified in the figure agree with the correct values $k_e = 10\; \text{s}^{-1}$, $k_r = 100\; \text{s}^{-1}$, and $k_{-} = 1\; \text{s}^{-1}$ of the numerical example within statistical errors. 

\begin{figure*}[t]
\begin{center}
\hspace*{-2cm}\resizebox{1.3\columnwidth}{!}{\includegraphics{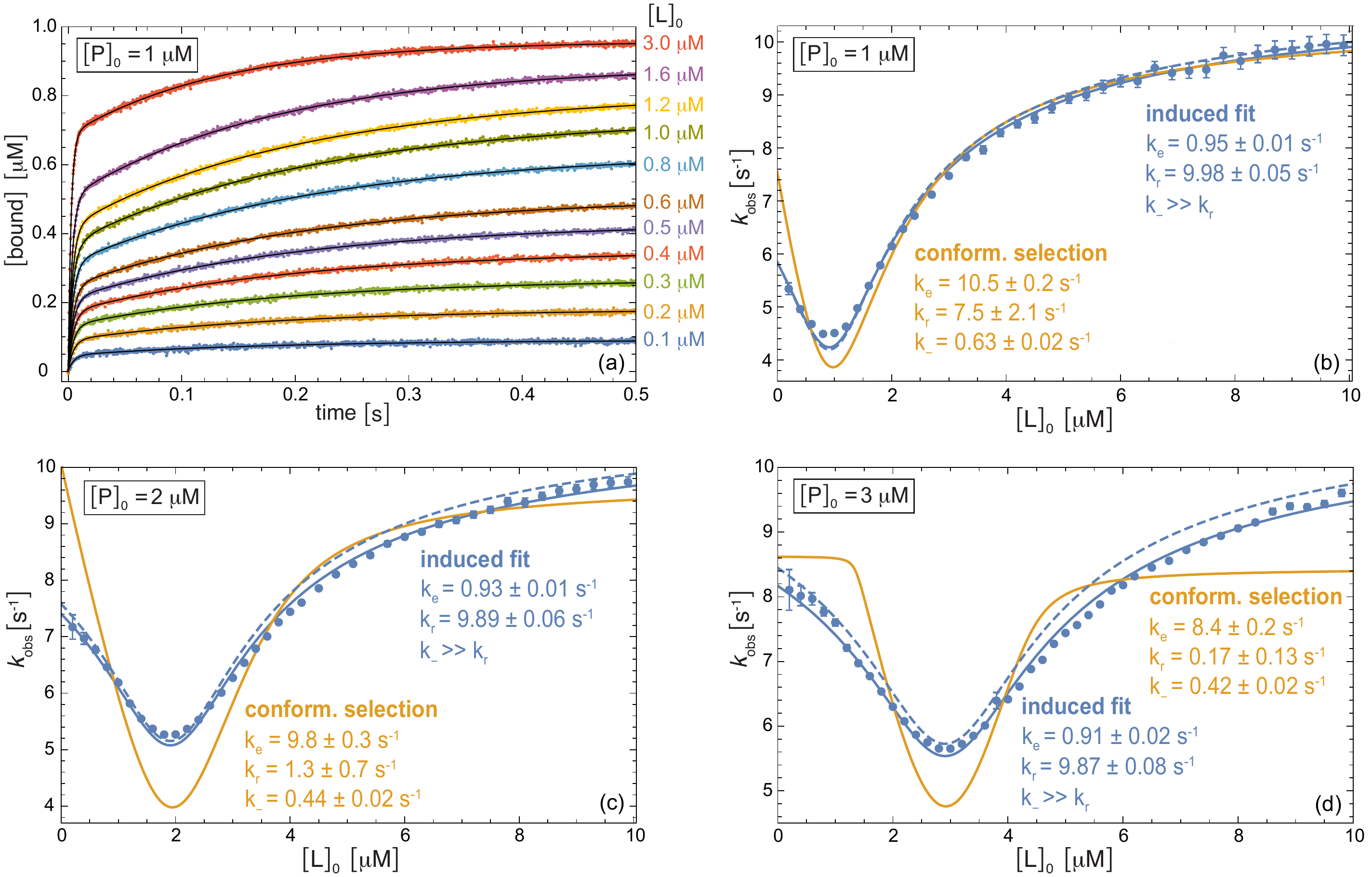}}
\end{center}
\caption{\small {\bf Numerical example for induced-fit binding} with the rate constants $k_{+} = 100 \; \upmu\text{M}^{-1} \text{s}^{-1}$, $k_{-} = 100\; \text{s}^{-1}$, $k_e = 1\; \text{s}^{-1}$, and $k_r = 10\; \text{s}^{-1}$. (a) Relaxation data for the bound complex obtained by numerical integration of the rate equations and subsequent addition of Gaussian noise with amplitude 0.004 $\upmu$M at the total protein concentration $[\text{P}]_0 = 1\; \upmu$M and exemplary total ligand concentrations $[\text{L}]_0$. The black lines represent multi-exponential fits of the data points. (b) to (d) Comparison of $k_\text{obs}$ values obtained from multi-exponential fits of numerical relaxation data (points) to our theoretical results for $k_\text{obs}$ (lines) at the three different total protein concentrations $[\text{P}]_0 = 1\, \upmu\text{M}$, $2\; \upmu\text{M}$, and $3\; \upmu\text{M}$ and total ligand concentrations $[\text{L}]_0$ between $0.1\, \upmu\text{M}$ and $10\, \upmu\text{M}$. The full lines represent fits of Eq.\ (\ref{if-kobs}) for induced-fit binding (blue) and of Eq.\ (\ref{cs-kobs}) for conformational-selection binding (orange), with fit parameter values specified in the figure. In these fits, the dissociation constant $K_d = 1/11 \; \upmu\text{M}$ is assumed to be known from equilibrium data. The dashed blue lines are obtained from Eq.\ (\ref{if-kobs}) for the `true' rate constants of the numerical example. 
}
\label{figure-example-IF}
\end{figure*}

In Fig.\ \ref{figure-example-IF}, we consider an induced-fit binding process with rate constants $k_{+} = 100 \; \upmu\text{M}^{-1} \text{s}^{-1}$, $k_{-} = 100\; \text{s}^{-1}$, $k_e = 1\; \text{s}^{-1}$, and $k_r = 10\; \text{s}^{-1}$ as a second numerical example. The $k_\text{obs}$ data points in Figs.\ \ref{figure-example-IF}(b) to (d) are again obtained from multi-exponential fits of relaxation curves that have been generated by numerical integration of the rate equations and  subsequent addition of Gaussian noise (see Fig.\ \ref{figure-example-IF}(a)).  The fits in Fig.\ \ref{figure-example-IF}(b) to (d)   
clearly identify induced-fit binding as the correct mechanism in this example. The full blue curves that represent fits of Eq.\ (\ref{if-kobs}) for induced-fit binding are in overall agreement with the $k_\text{obs}$ points, while the orange fit curves of Eq.\ (\ref{cs-kobs}) for conformational-selection binding deviate from the data. The fit values of the conformational transition rates $k_e$ and $k_r$ for the induced-fit binding model are in good agreement with the correct values $k_e = 1\; \text{s}^{-1}$, and $k_r = 10\; \text{s}^{-1}$ of the example. The dashed blue curves in Figs.\ \ref{figure-example-IF}(b) to (d), which are obtained from Eq.\ (\ref{if-kobs}) for the `true' rate constants of the induced-fit binding process, are in overall agreement with the data points. Slight deviations result from the fact that the amplitude of the slow relaxation mode with rate $k_\text{obs}$ is rather small compared to the amplitude of the fast modes (see Fig.\ \ref{figure-example-IF}(a)), which can lead to numerical inaccuracies.

In both numerical examples of Figs.\ \ref{figure-example-CS} and \ref{figure-example-IF}, the correct binding mechanism cannot be identified under pseudo-first-order conditions  because $k_\text{obs}$ is monotonically increasing with $[\mathrm{L}]_0$ for ligand concentrations that greatly exceed the protein concentration $[\mathrm{P}]_0$ \cite{Vogt12}.

\subsection*{Analysis of chemical relaxation rates for recoverin binding}

Chakrabarti et al.\ \cite{Chakrabarti16} have recently investigated the conformational dynamics and binding kinetics of the protein recoverin with chemical relaxation and advanced NMR experiments. Recoverin exhibits a  conformational change during binding of its ligand, which is a  rhodopsin kinase peptide fused to the B1 domain of immunoglobulin protein G in the experiments of Chakrabarti et al.\ \cite{Chakrabarti16}. The data points in Fig.\ 4 represent the dominant relaxation rates $k_\text{obs}$ obtained by Chakrabarti et al.\  from relaxation experiments at the temperatures 30\degree C and 10\degree C for a recoverin concentration of $10 \, \micro\text{M}$.  The lines in Fig.\ 4 result from fitting our general results (\ref{if-kobs}) and (\ref{cs-kobs}) for the dominant relaxation rate $k_\text{obs}$ of induced-fit and conformational-selection binding processes. In these fits, we have used the values $K_d = 1.0 \pm 0.2\; \upmu\mathrm{M}$  and $K_d = 1.8 \pm 0.2\; \upmu\text{M}$ obtained by Chakrabarti et al.\  from isothermal titration calometry experiments at 30\degree C and 10\degree C, which reduces the parameters to $k_e$, $k_r$, and $k_{-}$. The fits of our general result  (\ref{cs-kobs}) for  conformational-selection binding are rather insensitive to the relaxation rate $k_r$, which is illustrated in Fig.\ 4 by nearly identical fits for $k_r = 100\; \text{s}^{-1}$ and  $k_r = 1000\; \text{s}^{-1}$ (see dashed and full blue lines). Our fit values for the conformational excitation rate $k_e$ specified in the figure caption agree with the values $k_e = 33 \pm 5\; \text{s}^{-1}$ and $k_e = 23 \pm 5\; \text{s}^{-1}$ obtained by Chakrabarti et al.\  from advanced NMR experiments at 30\degree C and 10\degree C, respectively. From these experiments, Chakrabarti et al.\  obtain the values $k_r = 990 \pm 100\; \text{s}^{-1}$ and $k_r = 920 \pm 200\; \text{s}^{-1}$ at 30\degree C and 10\degree C, which cannot be deduced from our fits of the $k_\text{obs}$ data because these fits are insensitive to $k_r$. The NMR experiments indicate that the higher-energy conformation of unbound recoverin resembles the ground-state conformation of bound recoverin \cite{Chakrabarti16} as required for the conformational-selection binding mechanism illustrated in Fig.\ 1(b), and that the excited-state conformation of unbound recoverin has the equilibrium occupancy $P_e = k_e/(k_r + k_e) = 3.2\%\pm 0.5\%$ at 30\degree C and $P_e = 2.4\% \pm 0.7\%$ at 10\degree C, relative to the ground-state conformation.

\begin{figure*}[t]
\begin{center}
\hspace*{-2cm}\resizebox{1.3\columnwidth}{!}{\includegraphics{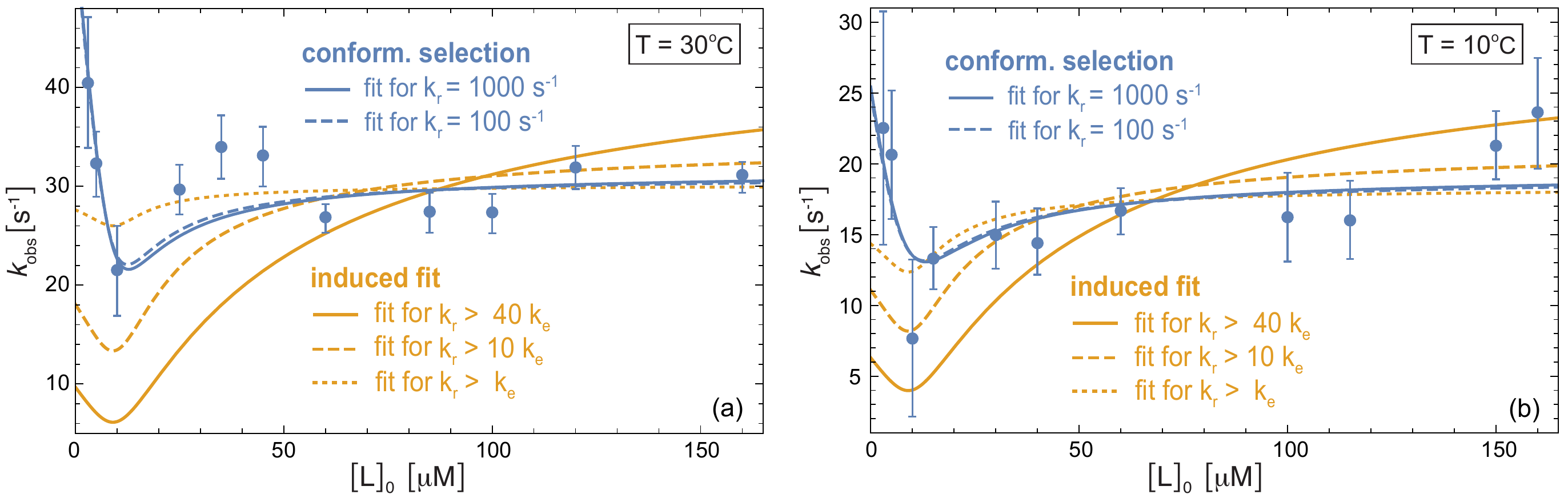}}
\end{center}
\caption{\small
{\bf Analysis of experimentally determined relaxation rates $k_\text{obs}$ for the binding of recoverin to a rhodopsin kinase peptide ligand}. The data points   
represent results of Chakrabarti et al.\ \cite{Chakrabarti16} obtained from chemical relaxation experiments at the temperatures 30\degree C and 10\degree C for a recoverin concentration of $10 \, \upmu\text{M}$.
The blue lines result from fits of Eq.\ (\ref{cs-kobs}) for conformational-selection binding with the values $k_r = 1000\; \text{s}^{-1}$ (full) and $k_r = 100\; \text{s}^{-1}$ (dashed) of the conformational relaxation rate. At 30\degree C, the parameter values obtained from fitting are $k_e = 31.5 \pm 0.8\; \text{s}^{-1}$ and $k_{-} = 5.1 \pm 0.4\; \text{s}^{-1}$ for $k_r = 1000\; \text{s}^{-1}$, and $k_e = 31.1 \pm 0.8\; \text{s}^{-1}$ and $k_{-} = 5.0 \pm 0.4\; \text{s}^{-1}$ for $k_r = 100\; \text{s}^{-1}$.
At 10\degree C, the fit parameter values are $k_e = 19.3 \pm 1.4\; \text{s}^{-1}$ and $k_{-} = 3.9 \pm 0.7\; \text{s}^{-1}$ for $k_r = 1000\; \text{s}^{-1}$, and $k_e = 19.0 \pm 1.3\; \text{s}^{-1}$ and $k_{-} = 3.8 \pm 0.7\; \text{s}^{-1}$ for $k_r = 100\; \text{s}^{-1}$.
The yellow lines represent fits of Eq.\ (\ref{if-kobs}) for induced-fit binding with constraints on the conformational excitation and relaxation rates $k_e$ and $k_r$. 
At 30\degree C, the obtained fit values for the conformational exchange rates  are $k_e = k_r = 15 \pm 10\; \text{s}^{-1}$ for the constraint $k_r > k_e$, $k_e = 3.1 \pm 1.9\; \text{s}^{-1}$ and $k_r = 31 \pm 4\; \text{s}^{-1}$  for the constraint $k_r > 10 k_e$, and $k_e = 1.1 \pm 0.8\; \text{s}^{-1}$ and $k_r = 44 \pm 8\; \text{s}^{-1}$  for $k_r > 40 k_e$. 
At 10\degree C, the fit values are $k_e = 4.5 \pm 4.0\; \text{s}^{-1}$ and $k_r = 14 \pm 10\; \text{s}^{-1}$ for the constraint $k_r > k_e$,
$k_e = 1.9 \pm 1.5\; \text{s}^{-1}$ and $k_r = 19 \pm 5\; \text{s}^{-1}$  for $k_r > 10 k_e$, and 
$k_e = 0.7 \pm 0.5\; \text{s}^{-1}$ and $k_r = 28 \pm 11\; \text{s}^{-1}$  for $k_r > 40 k_e$.
In all fits of Eq.\ (\ref{if-kobs}) for induced-fit binding, we obtain $k_{-}\gg k_r$, i.e.\ the fit values of the unbinding rate $k_{-}$ are much larger than the conformational relaxation rate $k_r$ and cannot be specified.
}
\label{figure-recoverin}
\end{figure*}

Fits of our general result  (\ref{if-kobs}) for the dominant relaxation rate $k_\text{obs}$ of induced-fit binding with unconstrained  parameters $k_e$, $k_r$, and $k_{-}$ lead to fit values for the conformational exchange rates $k_e$ and $k_r$ with $k_e\gg k_r$. For such values of $k_e$ and $k_r$, the conformation 1 of the induced-fit binding model illustrated in Fig.\ 1(a) is the ground-state conformation both for the unbound state and the bound state of recoverin, which contradicts the experimental observation that recoverin changes its conformation during binding \cite{Chakrabarti16}. Distinct ground-state conformations for the unbound and bound state of recoverin can be enforced by constraining $k_r$ to values larger than $k_e$. The yellow lines in Fig.\ 4 result from fits with the constraints $k_r> k_e$, $k_r> 10 k_e$, and $k_r> 40 k_e$. These constraints correspond to equilibrium occupancies $P_e$ of the excited-state conformation of bound recoverin with $P_e < 50\%$, $P_e < 9.1\%$, and $P_e < 2.4\%$, respectively. The fits of Eq.\ (\ref{if-kobs}) for induced-fit binding with the constraints $k_r> 10 k_e$ and $k_r> 40 k_e$ deviate rather strongly from the two data points with the smallest ligand concentrations $[\text{L}]_{0} = 3\, \upmu\text{M}$ and $5\, \upmu\text{M}$, in contrast to fits of Eq.\ (\ref{cs-kobs}) for conformational-selection binding (blue lines). A Bayesian model comparison of conformational-selection binding and induced-fit binding based on Eqs.\ (\ref{if-kobs}) and (\ref{cs-kobs}) leads to Bayes factors of $9.8\cdot 10^{13}$ and  $1.5 \cdot  10^{23}$ at $30^\circ$ C for the constraints $k_r> 10 k_e$ and $k_r> 40 k_e$, and to Bayes factors of $4.2  \cdot 10^{3}$ and  $9.6  \cdot 10^{9}$  at  10\degree C for  $k_r> 10 k_e$, and $k_r> 40 k_e$, respectively (see Methods for details). These Bayes factors indicate that the $k_\text{obs}$ data of Fig.\ 4 strongly point towards conformational-selection binding. Bayes factors larger than $10^2$ are generally considered to be decisive \cite{Jarosz14}. For the bound recoverin complex, Chakrabarti et al.\ did not observe an excited-state conformation in NMR experiments, which limits the excited-state occupancy $P_e$ to undetectable values smaller than $1\%$ for a conformational exchange that is fast compared to the NMR timescale as in the case of unbound recoverin. The analysis of the experimental data for the dominant relaxation rate $k_\text{obs}$ of recoverin binding based on our general results (\ref{if-kobs}) and (\ref{cs-kobs}) thus indicates a conformational-selection binding mechanism, in agreement with a numerical analysis of Chakrabarti et al.\ \cite{Chakrabarti16}. In this numerical analysis, Chakrabarti et al.\ include the chemical relaxation data for recoverin binding, additional relaxation data from dilution experiments, the values for the conformational exchange rates $k_e$ and $k_r$ obtained from NMR experiments, and  the $K_d$ values deduced from isothermal titration calometry \cite{Chakrabarti16}. In contrast, our analysis of the $k_\text{obs}$ data in Fig.\ 4 from the chemical relaxation experiments of recoverin binding only includes the $K_d$ values from isothermal titration calometry as additional input.

\section*{Discussion and conclusions}

We have shown here that the dominant rate $k_\text{obs}$ of chemical relaxation experiments with total protein and ligand concentrations of comparable magnitude conveys information on the binding mechanism and conformational transition rates of proteins. 
For sufficiently large protein concentrations $[\text{P}]_0$, the function $k_\text{obs}([\text{L}]_0)$ obtained from such experiments has characteristic features that are clearly distinct for induced-fit binding and conformational-selection binding.
The function $k_\text{obs}([\text{L}]_0)$ of induced-fit binding exhibits a characteristic symmetry around a minimum and tends to identical values for small and large ligand concentrations  $[\text{L}]_0$ as in Fig.\ 1(c) if the protein concentration $[\text{P}]_0$, which determines the location of the minimum, is sufficiently large. In contrast, the function $k_\text{obs} ([\text{L}]_0)$ of conformational-selection binding is either monotonically decreasing for $k_e<k_{-}$, or asymmetric around a minimum for $k_e>k_{-}$. In both cases, $k_\text{obs} ([\text{L}]_0)$ tends for small ligand concentrations $[\text{L}]_0$ to values that exceed the values for large ligand concentrations as in  Fig.\ 1(d) and (e) if the protein concentration $[\text{P}]_0$ is sufficiently large.

Our general results for the dominant rate $k_\text{obs}$ of chemical relaxation experiments thus provide a transparent route to distinguish induced-fit binding from conformational-selection binding based on the shape of the function $k_\text{obs}([\text{L}]_0)$, and to infer conformational transition rates from fitting. Alternatively, these binding mechanisms can be identified from a numerical analysis of time-dependent relaxation curves \cite{Daniels14,Daniels15,Chakrabarti16}, based on steric effects that may prohibit ligand entry and exit in the bound ground-state conformation of the protein and, thus, rule out conformational-selection binding \cite{Sullivan08}, from a comparison of conformational excitation rates to overall, effective binding and unbinding rates \cite{Boehr06a,Weikl14}, or from the effect of distal mutations that mainly affect the conformational exchange, but not the binding kinetics in different protein conformations \cite{Weikl14,Weikl09,Weikl12,Peuker13}.  Of particular interest is how such mutations change the overall binding and unbinding rates. If both conformational-selection and induced-fit binding are viable, increasing the ligand concentration may shift the binding mechanism from conformational selection to induced fit \cite{Weikl09,Hammes09,Daniels14,Greives14,Suddala15}.

\section*{Methods}

\subsection*{Near-equilibrium relaxation of induced-fit binding}

The induced-fit binding model of Fig.~\ref{figure_general}(a) leads to the four rate equations
\begin{align}
&\ddt[\mathrm{P_1}]  =  -k_{+}[\mathrm{P_1}][\mathrm{L}]+k_{-}[\mathrm{P_1L}] \\
&\ddt[\mathrm{L}]  =  -k_{+}[\mathrm{P_1}][\mathrm{L}]+k_{-}[\mathrm{P_1L}] \\
&\ddt[\mathrm{P_1L}]  =  k_{+}[\mathrm{P_1}][\mathrm{L}]-k_{-}[\mathrm{P_1L}]+k_{e}[\mathrm{P_2L}]-k_{r}[\mathrm{P_1L}] \\
&\ddt[\mathrm{P_2L}]  =  k_{r}[\mathrm{P_1L}]-k_{e}[\mathrm{P_2L}]
\end{align}
that describe the time-dependent evolution of the concentration $[\mathrm{P_1}]$ of the unbound protein, the concentration $[\mathrm{L}]$ of the unbound ligand, and the concentrations $[\mathrm{P_1L}]$ and  $[\mathrm{P_2L}]$ of the bound complexes.
These four rate equations are redundant because the total concentrations $[\mathrm{P}]_{0}$ and $[\mathrm{L}]_{0}$ of proteins and ligands are conserved:
\begin{align}
[\mathrm{P_1L}]+[\mathrm{P_2L}]+[\mathrm{P_1}]=[\mathrm{P}]_{0} 
\label{if-cons1} \\
[\mathrm{L}] + [\mathrm{P_1L}] + [\mathrm{P_2L}] = [\mathrm{L}]_0
\label{if-cons2}
\end{align}
With Eqs.\ (\ref{if-cons1}) and (\ref{if-cons2}), the concentrations $[\mathrm{P_1}]$ and $[\mathrm{P_1L}]$ can be expressed in terms of $[\mathrm{L}]$ and $[\mathrm{P_2L}]$, which results in the two non-redundant rate equations
\begin{align}
&\ddt[\mathrm{L}]  =   -k_{+}([\mathrm{L}]-[\mathrm{L}]_{0}+[\mathrm{P}]_{0})[\mathrm{L}]  + k_{-}([\mathrm{L}]_{0}-[\mathrm{L}]-[\mathrm{P_2L}]) 
\label{if-rate-eq-1}\\
&\ddt[\mathrm{P_2L}]   =   k_{r}([\mathrm{L}]_{0}-[\mathrm{L}]-[\mathrm{P_2L}])-k_{e}[\mathrm{P_2L}] 
\label{if-rate-eq-2}
\end{align}
These rate equations can be written in the vectorial form
\begin{equation}
\ddt\mathbf{c}=\mathbf{F}(\mathbf{c})
\label{if-vector-ode}
\end{equation}
with
\begin{equation}
\mathbf{c}(t)\equiv
\begin{pmatrix}
   [\mathrm{L}](t)\\
   [\mathrm{P_2L}](t)
\end{pmatrix}
\end{equation}
The two components of the vector $\mathbf{F}(\mathbf{c})$ in Eq.\ (\ref{if-vector-ode}) are the right-hand sides of the Eqs.\ (\ref{if-rate-eq-1}) and (\ref{if-rate-eq-2}). The rate equations describe the temporal evolution of the concentrations  $[\mathrm{L}]$ and $[\mathrm{P_2L}]$ towards equilibrium, and are nonlinear because of the quadratic term in $[\mathrm{L}]$ on the right-hand side of Eq.\ (\ref{if-rate-eq-1}).

To obtain linearized versions of the rate equations that describe the slow processes corresponding to the final relaxation into equilibrium, we expand the vector $\mathbf{F}(\mathbf{c})$ in Eq.\ (\ref{if-vector-ode}) around the equilibrium concentrations
$\mathbf{c}_{\mathrm{eq}}$:
\begin{equation}
\mathbf{F}(\mathbf{c})=\mathbf{F}(\mathbf{c}_{\mathrm{eq}}+\Delta\mathbf{c})\simeq\mathbf{F}(\mathbf{c}_{\mathrm{eq}})+J(\mathbf{c}_{\mathrm{eq}})\Delta\mathbf{\mathbf{c}}=J(\mathbf{c}_{\mathrm{eq}})\Delta\mathbf{\mathbf{c}}\label{eq:if-taylor}
\end{equation}
Here, $J$ is the Jacobian matrix of $\mathbf{F}$ with elements $J_{ij}=\partial F_{i}/\partial c_{j}$. The right-hand side of Eq.\ (\ref{eq:if-taylor}) follows from $\mathbf{F}(\mathbf{c}_{\mathrm{eq}})=0$. Inserting the expansion (\ref{eq:if-taylor}) into Eq.\  (\ref{if-vector-ode}) and making use of
$\tfrac{\mathrm{d}}{\mathrm{d}t}\mathbf{c}=\tfrac{\mathrm{d}}{\mathrm{d}t}(\mathbf{c}_{\mathrm{eq}}+\Delta\mathbf{c})=\tfrac{\mathrm{d}}{\mathrm{d}t}\Delta\mathbf{c}$ leads to the linearized rate equations
\begin{equation}
\ddt\Delta\mathbf{c}=J(\mathbf{c}_{\mathrm{eq}})\Delta\mathbf{\mathbf{c}}
\label{if-linear}
\end{equation}
with
\begin{equation}
J(\mathbf{c}_{\mathrm{eq}})=
\begin{pmatrix}
k_{+}\left([\mathrm{L}]_{0}-2[\mathrm{L}]_{\mathrm{eq}}-[\mathrm{P}]_{0}\right)-k_{-} & -k_{-}\\
-k_{r} & -k_{e}-k_{r}
\end{pmatrix}
\end{equation}
and the equilibrium concentration of the unbound ligand 
\begin{equation}
[\mathrm{L}]_{\mathrm{eq}}=  \frac{1}{2} \Big([\mathrm{L}]_{0} - [\mathrm{P}]_{0} - K_d   +\sqrt{\left([\mathrm{L}]_{0}-[\mathrm{P}]_{0}+K_d\right)^{2}+4[\mathrm{P}]_{0}K_d}\Big)
\label{if-Leq}
\end{equation}
The overall dissociation constant $K_d$ of the induced-fit binding process is given in Eq.\ (\ref{if-Kd}). The relaxation rates of the linearized rate equations (\ref{if-linear}) are the two eigenvalues of the matrix $-J(\mathbf{c}_{\mathrm{eq}})$. These eigenvalues are $k_\text{obs}$ given in Eq.\ (\ref{if-kobs}) and 
\begin{equation}
k_2=k_e+k_r+\frac{1}{2}\gamma + \frac{1}{2}\sqrt{\gamma^{2} + 4k_{-}k_r }
\end{equation}
with $\gamma$ and $\delta$ given in Eqs.\  (\ref{if-gamma}) and (\ref{delta}). The relaxation rate  $k_\text{obs}$ is smaller than $k_2$ and, thus, dominates the final relaxation into equilibrium. 

\subsection*{Near-equilibrium relaxation of conformational-selection binding}
\label{sec:CS-methods}

The four rate equations of the conformational-selection binding model of Fig.~\ref{figure_general}(b) are
\begin{align}
&\ddt[\mathrm{P_1}] =  -k_{e}[\mathrm{P_1}]+k_{r}[\mathrm{P_2}]\label{eq:cs-rate-e1}\\
&\ddt[\mathrm{P_2}]  =  k_{e}[\mathrm{P_1}]-k_{r}[\mathrm{P_2}]+k_{-}[\mathrm{P_2L}]-k_{+}[\mathrm{P_2}][\mathrm{L}]\label{eq:cs-rate-e2}\\
&\ddt[\mathrm{L}]  =  k_{-}[\mathrm{P_2L}]-k_{+}[\mathrm{P_2}][\mathrm{L}]\label{eq:cs-rate-l}\\
&\ddt[\mathrm{P_2L}]  =  -k_{-}[\mathrm{P_2L}]+k_{+}[\mathrm{P_2}][\mathrm{L}]\label{eq:cs-rate-b2}
\end{align}
The total concentrations $[\text{L}]_0$ and $[\text{P}]_0$ of the ligands and proteins are conserved:
\begin{align}
[\mathrm{L}] + [\mathrm{P_2L}]  =  [\mathrm{L}]_{0}
\label{cs-cons1} \\
[\mathrm{P_1}]+[\mathrm{P_2}]+[\mathrm{P_2L}]  =  [\mathrm{P}]_{0}
\label{cs-cons2}
\end{align}
With these equations, the concentrations $[\mathrm{P_1}]$ and $[\mathrm{P_2L}]$ can be expressed in terms of $[\mathrm{L}]$ and $[\mathrm{P_2}]$, which leads to the two  rate equations
\begin{align}
&\ddt[\mathrm{P_2}]  = k_e\left([\text{P}]_0-[\text{P}_2]\right)-(k_r + k_{+}[\mathrm{L}])[\text{P}_2]+(k_{-}-k_{e})([\mathrm{L}]_{0}-[\mathrm{L}])
\label{cs-rate-eq-1}\\
&\ddt[\mathrm{L}]  =  k_{-}([\mathrm{L}]_{0}-[\mathrm{L}])-k_{+}[\mathrm{P_2}][\mathrm{L}]
\label{cs-rate-eq-2}
\end{align}
These rate equations can be written in the vectorial form of Eq.\ (\ref{if-vector-ode}) with 
\begin{equation}
\mathbf{c}(t)\equiv
\begin{pmatrix}
   [\mathrm{P_2}](t)\\
   [\mathrm{L}](t)
\end{pmatrix}
\end{equation}
and with a vector $\mathbf{F}(\mathbf{c})$ that contains the right-hand sides of the Eqs.\ (\ref{cs-rate-eq-1}) and (\ref{cs-rate-eq-2}) as components. An expansion of the vector $\mathbf{F}(\mathbf{c})$ around the equilibrium concentrations $\mathbf{c}_{\mathrm{eq}}$ leads to Eq.\ (\ref{if-linear}) with the Jacobian matrix 
\begin{align}
J(\mathbf{c}_{\mathrm{eq}})=-\begin{pmatrix}k_{r}+k_{e}+k_{+}[\mathrm{L}]_{\mathrm{eq}} & -k_{e}+k_{-}+k_{+}[\mathrm{P_2}]_{\mathrm{eq}}\\
k_{+}[\mathrm{L}]_{\mathrm{eq}} & k_{-}+k_{+}[\mathrm{P_2}]_{\mathrm{eq}}
\end{pmatrix}\label{eq:cs-jaco}
\end{align}
and the equilibrium concentrations 
\begin{align}
&[\mathrm{P_2}]_{\mathrm{eq}}  =  \frac{1}{2K_d}\frac{k_{-}}{k_{+}}\Big([\mathrm{P}]_{0}-[\mathrm{L}]_{0}-K_d +\sqrt{([\mathrm{P}]_{0}-[\mathrm{L}]_{0}-K_d)^{2}+4K_d[\mathrm{P}]_{0}}\Big)\\
& [\mathrm{L}]_{\mathrm{eq}} =  \frac{1}{2}\left([\mathrm{L}]_{0} -[\mathrm{P}]_{0} -K_d +\sqrt{([\mathrm{P}]_{0}-[\mathrm{L}]_{0}-K_d)^{2}+4K_d[\mathrm{P}]_{0}}\right)
\label{CS-Leq}
\end{align}
The overall dissociation constant $K_d$ of the conformational-selection  binding process is given in Eq.\ (\ref{cs-Kd}). The relaxation rates of the linearized rate equations are the two eigenvalues of the matrix $-J(\mathbf{c}_{\mathrm{eq}})$. These eigenvalues are $k_\text{obs}$ given in Eq.\ (\ref{cs-kobs}) and 
\begin{equation}
k_2=k_{e}+\frac{1}{2}\alpha + \frac{1}{2}\sqrt{\alpha^{2}+\beta}\label{eq:cs-eigenvalues}
\end{equation}
with $\alpha$ and $\beta$ given in Eqs.\  (\ref{cs-alpha}) and (\ref{cs-beta}). The relaxation rate  $k_\text{obs}$ is smaller than $k_2$ and therefore dominates the final relaxation into equilibrium. 

To derive Eq.\ (\ref{L0min_CS}) for the location of the minimum of $k_\text{obs}$ as a function of the total ligand concentration $[\text{L}]_0$, we now consider the near-equilibrium relaxation of the conformational-selection model in quasi-steady-state approximation (qssa), which assumes that the concentration of the intermediate $[\text{P}_2]$ does not change in time. The left-hand side of Eq.\ (\ref{cs-rate-eq-1}) then is equal to zero, and the two Eqs.\ (\ref{cs-rate-eq-1}) and (\ref{cs-rate-eq-2}) reduce to the single equation
\begin{equation}
\ddt[\mathrm{L}] = - k_e k_{-}\frac{ \left( [\mathrm{L}] +  K_d\right)\left([\mathrm{L}] - [\mathrm{L}]_{0}\right) +  [\mathrm{L}][\mathrm{P}]_{0}}{k_{-}[\mathrm{L}] + k_eK_d} = f( [\mathrm{L}])
\end{equation}
An expansion of  the function $f( [\mathrm{L}])$ around the equilibrium concentration $[\mathrm{L}]_\text{eq}$ leads to the linear equation 
$\mathrm{d}[\mathrm{L}] /\mathrm{d}t\simeq - k_\text{obs}^\text{(qssa)} \left([\mathrm{L}] - [\mathrm{L}]_\text{eq}\right)$ with
\begin{align}
k_\text{obs}^\text{(qssa)}  =- \frac{\mathrm{d}f([\mathrm{L}])}{\mathrm{d}[\mathrm{L}]}\bigg|_{[\mathrm{L}]=[\mathrm{L}]_\text{eq}} =
\frac{k_{-} k_e \delta}{k_e K_d  + k_{-} [\mathrm{L}]_\text{eq}} 
\end{align}
and $\delta$ and $[\mathrm{L}]_\text{eq}$ given in Eqs.\ (\ref{delta}) and (\ref{CS-Leq}). The derivative of $k_\text{obs}^\text{(qssa)}$ is zero at $[\mathrm{L}]_0 = [\mathrm{L}]_0^\text{min}$ with $[\mathrm{L}]_0^\text{min}$ given in Eq.\ (\ref{L0min_CS}). In general, the quasi-steady-state result $k_\text{obs}^\text{(qssa)}$ is a good approximation of $k_\text{obs}$ if the rates for the transitions out of the intermediate state $\text{P}_2$ of conformational-selection binding are much larger than the rates for the transitions to $\text{P}_2$. A numerical analysis shows that the location $[\mathrm{L}]_0^\text{min}$ of the minimum of $k_\text{obs}^\text{(qssa)}([\mathrm{L}])$ is in good agreement with the location of the minimum of $k_\text{obs}([\mathrm{L}])$ for conformational transitions rates with $k_r\gg k_e$.

\subsection*{Multi-exponential relaxation} 

In the numerical examples illustrated in Figs.\ \ref{figure-example-CS} and \ref{figure-example-IF}, chemical relaxation curves for conformational-selection and induced-fit binding are fitted with a multi-exponential model. Such multi-exponential models are an adequate description for the time evolution of concentrations in first-order chemical reactions. However, the binding steps of the induced-fit and conformational-selection models of Fig.\ \ref{figure_general}(a) and (b) are of second order. To justify that multi-exponential models can also be used to approximate the chemical relaxation of second-order reactions, we consider here the elementary binding model
\begin{equation}
\ce{P + L <=>[k_{+}\lbrack\mathrm{P}\rbrack\lbrack\mathrm{L}\rbrack][k_{-}] PL}
\label{eb-scheme}
\end{equation}
of a protein P and ligand L. For the initial condition $[\mathrm{PL}](0)=0$, the rate equation of the elementary binding model can be written as
\begin{equation}
\frac{\mathrm{d}}{\mathrm{d}t}[\mathrm{PL}]=k_{+}\left([\mathrm{P}]_{0}-[\mathrm{PL}]\right)\left([\mathrm{L}]_{0}-[\mathrm{PL}]\right)-k_{-}[\mathrm{PL}]
\label{eb-rate-equation}
\end{equation}
 and has the analytical solution \cite{Peuker13}
\begin{equation}
[\mathrm{PL}](t) = - \frac{\lambda_1 \left(\mathrm{e}^{(\lambda_1 - \lambda_2)t}-1\right)}{k_\text{+}\left(\mathrm{e}^{(\lambda_1 - \lambda_2)t}-\lambda_1/\lambda_2\right)}
\label{eb-analytical-solution}
\end{equation}
with
\begin{equation}
\lambda_{1,2} = -\frac{1}{2}k_{+}\left([\mathrm{P}]_0 + [\mathrm{L}]_0 + K_d
\pm  \sqrt{\left([\mathrm{P}]_0 + [\mathrm{L}]_0 + K_d \right)^2 - 4 [\mathrm{P}]_0  [\mathrm{L}]_0} 
\right)
\end{equation}
where $K_d = k_{-}/k_{+}$ is the dissociation constant of the elementary binding model.

We first show that $\lambda_2 - \lambda_1$ is identical to the dominant relaxation rate $k_\text{obs}$ obtained from a linear expansion around equilibrium. An expansion of  the right-hand side of Eq.\ (\ref{eb-rate-equation}) around the equilibrium concentration 
\begin{equation}
[\mathrm{PL}]_{\mathrm{eq}}=\frac{1}{2}\left([\mathrm{P}]_{0}+[\mathrm{L}]_{0}+K_d 
- \sqrt{([\mathrm{L}]_{0}-[P]_{0}+K_{d})^{2}+4K_{d}[P]_{0}} 
\right)
\end{equation}
leads to the linear equation $\mathrm{d}[\mathrm{PL}] /\mathrm{d}t\simeq - k_\text{obs} \left([\mathrm{PL}] - [\mathrm{PL}]_\text{eq}\right)$ with
\begin{equation}
k_{\mathrm{obs}}=k_{+}\sqrt{([\mathrm{L}]_{0}-[P]_{0}+K_{d})^{2}+4K_{d}[P]_{0}}
\label{eb-kobs}
\end{equation}
This dominant relaxation rate $k_\text{obs}$ is identical to $\lambda_2 - \lambda_1$. As a function of $[\mathrm{L}]_{0}$, the dominant rate $k_\text{obs}$ of the elementary binding model exhibits a minimum at $[\mathrm{L}]_{0}^{\text{min}}= [P]_{0} - K_{d}$ and is symmetric with respect to this minimum.

We next use the limit of the geometric series $\sum_{n=0}^{\infty}q^{n}=1/(1-q)$
with $q=e^{-k_{\mathrm{obs}}t}\lambda_2/\lambda_1$ to rewrite Eq.\ (\ref{eb-analytical-solution}) as
\begin{equation}
[\mathrm{PL}](t)\propto\lambda_2+\left(\lambda_2- \lambda_1\right)\sum_{n=1}^{\infty}\frac{\mathrm{e}^{-n k_{\mathrm{obs}}t}}{\left(\lambda_1/\lambda_2\right)^{n}}
\label{eb-multi-exponential}
\end{equation}
which shows that the chemical relaxation of the elementary binding model can be described as an infinite sum of exponential functions. The exponents of these functions are integer multiples of $k_{\mathrm{obs}}$, which is reminiscent of the higher harmonics in oscillatory phenomena. The prefactors $(\lambda_2/\lambda_1)^{n}$ in Eq.\ (\ref{eb-multi-exponential}) decay exponentially with the order $n$ of the harmonic because of $\lambda_2/\lambda_1<1$. The infinite sum of Eq.\ (\ref{eb-multi-exponential}) therefore can be truncated in practical situations.  Under pseudo-first-order conditions, Eq.\ (\ref{eb-multi-exponential}) reduces to a single-exponential relaxation.

In analogy to the elementary binding model, we propose that the time evolution of the concentrations in the induced-fit and conformational-selection models can be represented as a sum of exponentials where the exponents are integer combinations $- i k_\text{obs} - j k_{2}$ with $i, j = 0, 1, 2, 3, ...$ of the relaxation rates $k_\text{obs}$ and $k_{2}$  obtained from a linear expansion around the equilibrium concentrations. Under pseudo-first-order conditions, the chemical relaxation reduces to a double-exponential relaxation \cite{Weikl09,Weikl12,Vogt12}.

In the numerical examples of Figs.\ \ref{figure-example-CS} and \ref{figure-example-IF}, the chemical relaxation of the bound complexes is fitted with a multi-exponential model
\begin{equation}
[\text{bound}](t) = A_{0} + \sum_{n=1}^{N}A_{n}e^{-k_{n}t}
\end{equation}
with $k_{n}>0$ for all $n$. We have used the routine NonlinearModelFit of the software Mathematica \cite{Mathematica} with the differential evolution algorithm \cite{Storn97}, which was repeatedly run with different values of its $F$ parameter ranging from 0.1 to 1 for a given number of exponentials $N$.  Among different runs, we have selected fit results based on the residual sum of squares, after discarding fits with singular results in which two rates $k_{n}$ coincide within 95\% confidence intervals, or in which one or more rates $k_{n}$ are identical to 0  within 95\% confidence intervals. We have then determined the number of exponentials $N$ based on the small-sample-size corrected version of Akaike's information criterion (AIC) \cite{Cavanaugh97}.

\subsection*{Bayes factors}

The Bayes factor $K$ is as measure for how plausible one model is relatively to an alternative model, given experimental data \cite{Jaynes03}. The Bayes factor for the plausibility of the  conformational-selection binding model relative to induced-fit binding model is
\begin{equation}
K=\frac{\int p(\text{data}\mid\text{conformational-selection binding},\theta)p(\theta)\mathrm{\mathrm{d}\theta}}{\int p(\text{data}\mid\text{induced-fit binding},\theta)p(\theta)\mathrm{\mathrm{d}\theta}}
\label{bayes}
\end{equation}
Here, $p(\text{data}\mid M,\theta)$ is the probability that the data were produced by the model $M$ with given parameters $\theta$, where $M$ either stands for conformational-selection binding or induced-fit binding, and $p(\theta)$ is the prior distribution on the parameter values, which encodes any prior knowledge that we have about the parameters. The integrals of Eq.\ (\ref{bayes}) are taken over all parameter values and result in the  probability  $p(\text{data}\mid M)$  that the data were produced by the model, regardless of specific parameter values. The data here consist of the slowest relaxation rates $k_{\mathrm{obs}}^{(i)}$ with $i = 1, 2, \ldots, N$ obtained from multi-exponential fits of the $N$ time series with ligand concentrations $[L]_0^{(i)}$, and the errors $\sigma_i$ of these rates. Following standard approaches \cite{Jaynes03}, the probability that the data were generated by the model $M$ with parameters $\theta=(k_{e},k_{r},k_{-},K_{d},[P]_{0})$ is
\begin{equation}
p(\text{data}\mid M,\theta)\propto
\prod_{i=1}^{N}\exp{\left[-\frac{\left(k_{\mathrm{obs}}^{(i)}-k_{\mathrm{obs}}^{M}\left(\theta,[L]_{0}^{(i)}\right)\right)^{2}}{2\sigma_{i}^{2}}\right]} 
\label{pMtheta}
\end{equation}
for $k_{r}>nk_{e}$, and 0 otherwise. The inequality $k_{r}>nk_{e}$ reflects constraints on the conformational relaxation rate $k_r$ and excitation rate $k_e$ of the models (see section ``Analysis of chemical relaxation rates for recoverin binding").  Eq.\ (\ref{pMtheta}) implies that the errors $k_{\mathrm{obs}}^{(i)}-k_{\mathrm{obs}}^{M}\left(\theta,[L]_{0}^{(i)}\right)$ are independently and normally distributed random variables with standard deviations $\sigma_i$. Depending on the model $M$, we either use Eq.\ (1) or (6) to determine $k_{\mathrm{obs}}^{M}\left(\theta,[L]_{0}^{(i)}\right)$. For simplicity, $K_{d}$ and $[P]_{0}$ are kept fixed at the experimentally measured values. We choose a prior $p(\theta)$ that is uniform in the logarithm of the rates $k_{e},k_{r},k_{-}$. Taking the logarithm of the rates is not crucial, as a uniform prior on the rates gives similar results in the analysis of recoverin binding and, thus, leads to the same conclusions. The prior $p(\theta)$ here can be chosen to be uniform 
because it is identical for both the induced-fit and conformational-selection binding models due to the equivalent parameters of the models \cite{Strachan05}.

\section*{Acknowledgments}
This research has been partially funded by Deutsche Forschungsgemeinschaft (DFG) through grant CRC 1114. 
F.\ P.\ would like to thank Christof Sch\"utte for insightful discussions.

\nolinenumbers


\small

\begin{thebibliography}{10}

\bibitem{Gerstein98}
Gerstein M, Krebs W.
\newblock A database of macromolecular motions.
\newblock Nucleic Acids Res. 1998;26:4280--4290.

\bibitem{Eisenmesser05}
Eisenmesser EZ, Millet O, Labeikovsky W, Korzhnev DM, Wolf-Watz M, Bosco DA,
  et~al.
\newblock Intrinsic dynamics of an enzyme underlies catalysis.
\newblock Nature. 2005;438:117--121.

\bibitem{Beach05}
Beach H, Cole R, Gill M, Loria J.
\newblock Conservation of $\mu$s-ms enzyme motions in the apo- and
  substrate-mimicked state.
\newblock J Am Chem Soc. 2005;127:9167--9176.

\bibitem{Boehr06a}
Boehr DD, McElheny D, Dyson HJ, Wright PE.
\newblock The dynamic energy landscape of dihydrofolate reductase catalysis.
\newblock Science. 2006;313:1638--1642.

\bibitem{Henzler07b}
Henzler-Wildman KA, Thai V, Lei M, Ott M, Wolf-Watz M, Fenn T, et~al.
\newblock Intrinsic motions along an enzymatic reaction trajectory.
\newblock Nature. 2007;450:838--844.

\bibitem{Tang07}
Tang C, Schwieters CD, Clore GM.
\newblock Open-to-closed transition in apo maltose-binding protein observed by
  paramagnetic {NMR}.
\newblock Nature. 2007;449:1078--1082.

\bibitem{Lange08}
Lange OF, Lakomek NA, Fares C, Schr\"oder GF, Walter KFA, Becker S, et~al.
\newblock {Recognition dynamics up to microseconds revealed from an RDC-derived
  ubiquitin ensemble in solution}.
\newblock Science. 2008;320:1471--1475.

\bibitem{Kim13}
Kim E, Lee S, Jeon A, Choi JM, Lee HS, Hohng S, et~al.
\newblock A single-molecule dissection of ligand binding to a protein with
  intrinsic dynamics.
\newblock Nat Chem Biol. 2013;9:313--318.

\bibitem{Munro14}
Munro JB, Gorman J, Ma X, Zhou Z, Arthos J, Burton DR, et~al.
\newblock Conformational dynamics of single {HIV-1} envelope trimers on the
  surface of native virions.
\newblock Science. 2014;346:759--763.

\bibitem{Ghoneim14}
Ghoneim M, Spies M.
\newblock Direct correlation of {DNA} binding and single protein domain motion
  via dual illumination fluorescence microscopy.
\newblock Nano Lett. 2014;14:5920--5931.

\bibitem{Ma99}
Ma B, Kumar S, Tsai CJ, Nussinov R.
\newblock Folding funnels and binding mechanisms.
\newblock Protein Eng. 1999;12:713--720.

\bibitem{Koshland58}
Koshland DE.
\newblock Application of a theory of enzyme specificity to protein synthesis.
\newblock Proc Natl Acad Sci USA. 1958;44:98--104.

\bibitem{Weikl14}
Weikl TR, Paul F.
\newblock Conformational selection in protein binding and function.
\newblock Protein Sci. 2014;23:1508--1518.

\bibitem{Bosshard01}
Bosshard HR.
\newblock {Molecular recognition by induced fit: How fit is the concept?}
\newblock News Physiol Sci. 2001;16:171--1733.

\bibitem{Sullivan08}
Sullivan SM, Holyoak T.
\newblock Enzymes with lid-gated active sites must operate by an induced fit
  mechanism instead of conformational selection.
\newblock Proc Natl Acad Sci USA. 2008;105:13829--13834.

\bibitem{Weikl09}
Weikl TR, von Deuster C.
\newblock Selected-fit versus induced-fit protein binding: kinetic differences
  and mutational analysis.
\newblock Proteins. 2009;75:104--110.

\bibitem{Boehr09}
Boehr DD, Nussinov R, Wright PE.
\newblock The role of dynamic conformational ensembles in biomolecular
  recognition.
\newblock Nat Chem Biol. 2009;5:789--796.

\bibitem{Hammes09}
Hammes GG, Chang YC, Oas TG.
\newblock Conformational selection or induced fit: a flux description of
  reaction mechanism.
\newblock Proc Natl Acad Sci USA. 2009;106:13737--13741.

\bibitem{Wlodarski09}
Wlodarski T, Zagrovic B.
\newblock Conformational selection and induced fit mechanism underlie
  specificity in noncovalent interactions with ubiquitin.
\newblock Proc Natl Acad Sci USA. 2009;106:19346--19351.

\bibitem{Changeux11b}
Changeux JP, Edelstein S.
\newblock Conformational selection or induced fit? 50 years of debate resolved.
\newblock F1000 Biol Rep. 2011;3:19.

\bibitem{Weikl12}
Weikl TR, Boehr DD.
\newblock Conformational selection and induced changes along the catalytic
  cycle of {{\em Escherichia coli}} dihydrofolate reductase.
\newblock Proteins. 2012;80:2369--2383.

\bibitem{Vogt12}
Vogt AD, Di~Cera E.
\newblock Conformational selection or induced fit? A critical appraisal of the
  kinetic mechanism.
\newblock Biochemistry. 2012;51:5894--5902.

\bibitem{Kiefhaber12}
Kiefhaber T, Bachmann A, Jensen KS.
\newblock Dynamics and mechanisms of coupled protein folding and binding
  reactions.
\newblock Curr Opion Struct Biol. 2012;22:21--29.

\bibitem{Vogt14}
Vogt AD, Pozzi N, Chen Z, Di~Cera E.
\newblock Essential role of conformational selection in ligand binding.
\newblock Biophys Chem. 2014;186:13--21.

\bibitem{Pozzi12}
Pozzi N, Vogt AD, Gohara DW, Di~Cera E.
\newblock Conformational selection in trypsin-like proteases.
\newblock Curr Opin Struct Biol. 2012;22:421--431.

\bibitem{Daniels14}
Daniels KG, Tonthat NK, McClure DR, Chang YC, Liu X, Schumacher MA, et~al.
\newblock Ligand concentration regulates the pathways of coupled protein
  folding and binding.
\newblock J Am Chem Soc. 2014;136:822--825.

\bibitem{Daniels15}
Daniels KG, Suo Y, Oas TG.
\newblock Conformational kinetics reveals affinities of protein conformational
  states.
\newblock Proc Natl Acad Sci USA. 2015;112:9352--9357.

\bibitem{Chakrabarti16}
Chakrabarti KS, Agafonov RV, Pontiggia F, Otten R, Higgins MK, Schertler GFX,
  et~al.
\newblock Conformational selection in a protein-protein interaction revealed by
  dynamic pathway analysis.
\newblock Cell Reports. 2016;14:32--42.

\bibitem{James03}
James LC, Roversi P, Tawfik DS.
\newblock Antibody multispecificity mediated by conformational diversity.
\newblock Science. 2003;299:1362--1367.

\bibitem{Heredia06}
Heredia VV, Thomson J, Nettleton D, Sun S.
\newblock Glucose-induced conformational changes in glucokinase mediate
  allosteric regulation: transient kinetic analysis.
\newblock Biochemistry. 2006;45:7553--7562.

\bibitem{Kim07}
Kim YB, Kalinowski SS, Marcinkeviciene J.
\newblock A pre-steady state analysis of ligand binding to human glucokinase:
  Evidence for a preexisting equilibrium.
\newblock Biochemistry. 2007;46:1423--1431.

\bibitem{Tummino08}
Tummino PJ, Copeland RA.
\newblock Residence time of receptor-ligand complexes and its effect on
  biological function.
\newblock Biochemistry. 2008;47:5481--5492.

\bibitem{Antoine09}
Antoine M, Boutin JA, Ferry G.
\newblock Binding kinetics of glucose and allosteric activators to human
  glucokinase reveal multiple conformational states.
\newblock Biochemistry. 2009;48:5466--5482.

\bibitem{Vogt13}
Vogt AD, Di~Cera E.
\newblock Conformational selection is a dominant mechanism of ligand binding.
\newblock Biochemistry. 2013;52:5723--5729.

\bibitem{Gianni14}
Gianni S, Dogan J, Jemth P.
\newblock Distinguishing induced fit from conformational selection.
\newblock Biophys Chem. 2014;189:33--39.

\bibitem{Vogt15}
Vogt AD, Chakraborty P, Di~Cera E.
\newblock Kinetic dissection of the pre-existing conformational equilibrium in
  the trypsin fold.
\newblock J Biol Chem. 2015;290:22435--22445.

\bibitem{Jarosz14}
Jarosz AF, Wiley J.
\newblock What are the odds? A practical guide to computing and reporting Bayes
  factors.
\newblock J Problem Solving. 2014;7:2.

\bibitem{Peuker13}
Peuker S, Cukkemane A, Held M, No{\'e} F, Kaupp UB, Seifert R.
\newblock Kinetics of ligand-receptor interaction reveals an induced-fit mode
  of binding in a cyclic nucleotide-activated protein.
\newblock Biophys J. 2013;104:63--74.

\bibitem{Greives14}
Greives N, Zhou HX.
\newblock Both protein dynamics and ligand concentration can shift the binding
  mechanism between conformational selection and induced fit.
\newblock Proc Natl Acad Sci USA. 2014;111:10197--10202.

\bibitem{Suddala15}
Suddala KC, Wang J, Hou Q, Walter NG.
\newblock Mg$^{2+}$ shifts ligand-mediated folding of a riboswitch from
  induced-fit to conformational selection.
\newblock J Am Chem Soc. 2015;137:14075--14083.

\bibitem{Mathematica}
Mathematica, Version 10.3.
\newblock Wolfram Research, Inc., Champaign, Illinois; 2015.

\bibitem{Storn97}
Storn R, Price K.
\newblock Differential evolution - a simple and efficient heuristic for global
  optimization over continuous spaces.
\newblock J Global Optim. 1997;11:341--359.

\bibitem{Cavanaugh97}
Cavanaugh JE.
\newblock Unifying the derivations of the {A}kaike and corrected {A}kaike
  information criteria.
\newblock Stat Probabil Lett. 1997;31:201--208.

\bibitem{Jaynes03}
Jaynes ET.
\newblock Probability theory: the logic of science.
\newblock Cambridge University Press; 2003.

\bibitem{Strachan05}
Strachan RW, van Dijk HK.
\newblock Improper priors with well-defined {B}ayes factors.
\newblock Liverpool, L69 7ZA, United Kingdom: Department of Economics and
  Accounting, University of Liverpool; 2005. EI 2004-18.

\end{thebibliography}

\clearpage

%
%
%

\end{document}